# Current and voltage based bit errors and their combined mitigation for the Kirchhoff-law–Johnson-noise secure key exchange

Yessica Saez · Laszlo B. Kish · Robert Mingesz · Zoltan Gingl · Claes G. Granqvist

**Abstract** We classify and analyze bit errors in the current measurement mode of the Kirchhoff-law–Johnson-noise (KLJN) key distribution. The error probability decays exponentially with increasing bit exchange period and fixed bandwidth, which is similar to the error probability decay in the voltage measurement mode. We also analyze the combination of voltage and current modes for error removal. In this combination method, the error probability is still an exponential function that decays with the duration of the bit exchange period, but it has superior fidelity to the former schemes.

**Keywords**

Information theoretic security · Second Law of Thermodynamics · Statistical physical key distribution · Unconditional security · Secure key distribution via wire

Y. Saez (✉) and L.B. Kish

Department of Electrical and Computer Engineering, Texas A&M University, College Station, TX 77843-3128, USA
e-mail: yessica.saez@neo.tamu.edu ,
laszlo.kish@ece.tamu.edu

R. Mingesz and Z. Gingl
Department of Technical Informatics, University of Szeged, Árpád tér 2, Szeged, H-6701, Hungary
e-mail: mingesz@inf.u-szeged.hu , gingl@inf.u-szeged.hu

C.G. Granqvist
Department of Engineering Sciences, The Ångström Laboratory, Uppsala University, P.O. Box 534, SE-75121 Uppsala, Sweden
e-mail: claes-goran.granqvist@angstrom.uu.se

## 1 Introduction

Information theoretic security, often referred to as "unconditional security" [1], means that the security measures are determined by information theory or, in physical systems, by measurement theory. These security measures can be perfect or imperfect and are determined by the eavesdropper's ("Eve's") supposed optimum conditions for extracting the maximum amount of information. In other words, Eve's information is calculated by assuming that she has unlimited computational power and that her measurement accuracy and measurement speed are limited only by the laws of physics and the protocol's conditions.

Quantum key distribution (QKD) [2] was the first scheme based on the laws of physics that claimed to possess unconditional security. However, this claim is not uncontested and there is an ongoing debate [3–7] about the security inherent in existing QKD schemes. This discussion was initiated by quantum security experts Horace Yuen [3–4, 7] and Osamu Hirota [5], who agreed in their claim that the achievable level of security in QKD schemes is questionable. Renner [6] later entered this debate to defend the foundations of quantum cryptography and to validate existing security proofs.

From a practical point of view one observes that several communicators, including commercial and laboratory-type QKD devices, have been successfully cracked, as shown in numerous publications [8–22]. These demonstrated flaws of the QKD devices—and also some practical issues such as limited communication range and high price—have inspired new initiatives that involve non-QKD schemes utilizing alternative types of mechanisms to achieve security [23, 24].

Recent studies have shown that a system employing two pairs of resistors, with Gaussian voltage noise generators to imitate and enhance their Johnson noise, can be used for secure key distribution [25–30]. This system is known as the Kirchhoff-law–Johnson-noise (KLJN) secure key distribution and provides information theoretic security [26, 27]. It is based on Kirchhoff's loop law of quasi-electrodynamics and the fluctuation–dissipation theorem of statistical physics [25–31]. The KLJN scheme has potential applications including physical uncloneable function hardware keys [32]; unconditional security within computers, hardware and other instruments [32, 33]; and unconditionally secure smart grids [34–36].

Figure 1 shows the fundamental KLJN system [25–30] without defense elements against active (invasive) attacks and vulnerabilities represented by non-ideal building elements. Under practical conditions, this system utilizes enhanced Johnson noise with high noise temperature, obtained from Gaussian noises generated electronically so that quasi-static and thermodynamic characteristics are emulated as accurately as possible, in order to approach perfect security [31]. The core KLJN channel is represented by a wire line to which the two communicating parties, "Alice" and "Bob", connect their resistors $R_A$ and $R_B$, respectively. These resistors are randomly selected from the set $\{R_0, R_1\}$, with $R_0 \neq R_1$. The resistor $R_0$ indicates the low (0) bit and the resistor $R_1$ indicates the high (1) bit, respectively [25]. At the beginning of each clock period or bit exchange period, Alice and Bob, who have identical pairs of resistors, randomly choose one of these resistors and connect it to the wire line. The Gaussian voltage noise generators represent either the Johnson noises of the resistors or external noise generators delivering band-limited white noise with publicly known bandwidth and effective noise temperature $T_{eff}$ [25–26, 30]. According to the fluctuation–dissipation theorem, the enhanced Johnson noise voltages of Alice's and Bob's resistors—denoted $u_A(t)$ and $u_B(t)$ respectively, where $u_A \in \{u_{0,A}(t), u_{1,A}(t)\}$ and $u_B \in \{u_{0,B}(t), u_{1,B}(t)\}$—generate a channel noise voltage $u_c(t)$ between the wire line and ground as well as a channel noise current $i_c(t)$ in the wire.

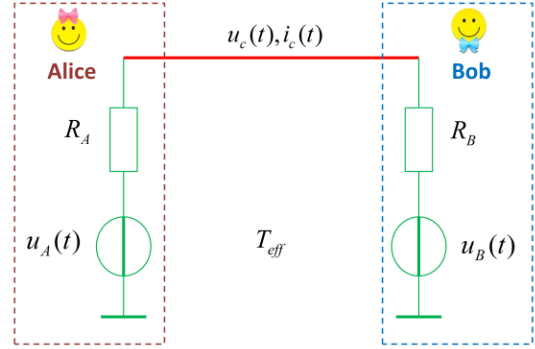

**Fig. 1** Outline of the core KLJN secure key exchange scheme without defense circuitry (current/voltage monitoring/comparison) against invasive attacks or attacks utilizing non-ideal components and conditions. $T_{eff}$ is the effective noise temperature, $R_A$, $u_A(t)$, $R_B$, and $u_B(t)$ are the resistor values and noise voltages at Alice and Bob, respectively. $u_c(t)$ and $i_c(t)$ are channel noise voltage and current, respectively.

Within the bit exchange period, Alice and Bob measure the mean-square channel noise voltage and/or current amplitudes $\langle u_c^2(t) \rangle$ and/or $\langle i_c^2(t) \rangle$. By applying Johnson's noise formula and Kirchhoff's loop law, it follows that the theoretical values of the mean-square noise voltage and current for a given channel noise bandwidth $B_{KLJN}$ and temperature $T_{eff}$ are [25, 26]

$$\langle u_c^2(t) \rangle = S_{u,c}(f) B_{KLJN} = 4kT_{eff} R_{\|} B_{KLJN} \ , \quad (1a)$$

$$\langle i_c^2(t) \rangle = S_{i,c}(f) B_{KLJN} = 4kT_{eff} \frac{1}{R_{loop}} B_{KLJN} \ . \quad (1b)$$

Here $\langle \ \rangle$ represents ideal infinite-time average, $S_{u,c}(f)$ is the power density spectrum of the channel voltage noise, $S_{i,c}(f)$ is the power density spectrum of channel current noise, $k$ is Boltzmann's constant,
2

$R_{\parallel} = R_A R_B / (R_A + R_B)$, and $R_{loop} = R_A + R_B$.

The resistance values $R_{\parallel}$ and/or $R_{loop}$ can be publicly known by comparing the result of the measurement of the mean-square channel noise voltage and/or current amplitudes with the corresponding theoretical values obtained from Eq. (1). Alice and Bob know their own chosen resistors, and hence the total resistances $R_{\parallel}$ and/or $R_{loop}$ allow them to deduce the resistance value and actual bit status at the other end of the wire.

The cases when Alice and Bob use the same resistance values—*i.e.*, the 00 and 11 situations—represent *non-secure* bit exchange. Eve will then be able to find the resistor values, their location and the status of the bits, because the total resistance will either be the lowest or the highest value of the three possible magnitudes of the total resistance. The situations when Alice and Bob use the resistance values 01 and 10 signify a *secure* bit exchange event because these resistances cannot be distinguished by measured mean-square values. Alice and Bob will know that the other party has the inverse of his/her bit, which implies that a secure key exchange takes place.

The KLJN key distribution scheme has statistical errors due to the finite duration time $\tau$ of the bit-exchange period [30, 31]. Specifically, an experimental demonstration of the KLJN scheme, conducted recently by Mingesz *et al.* [30], yielded that the fidelity of the KLJN key exchange was 99.98 %, corresponding to a bit error probability of 0.02 %.

The bit errors were analyzed recently by Saez and Kish [31] for the case of the mean square noise *voltage* being utilized for key exchange. The bit error probability showed exponential decay vs. $\tau$. In the present paper we analyze the bit errors in the *current* measurement mode, and we also analyze the *combination* of voltage and current modes for error mitigation.

## 2 Bit interpretation of the measured channel current

We suppose ideal components/conditions and proceed as in earlier work [31]. Alice and Bob obtain the total loop resistance by measuring the mean-square channel noise current amplitude $\langle i_c^2(t) \rangle_{\tau}$, where $\langle \ \rangle_{\tau}$ indicates a finite-time average over random fluctuations around the exact mean-square noise current. Figure 2 illustrates the three possible levels of the measured mean-square channel noise current. The 11, 01/10 and 00 bit situations result in mean-square channel noise currents $\langle i_{11}^2(t) \rangle_{\tau}$, $\langle i_{01/10}^2(t) \rangle_{\tau}$ and $\langle i_{00}^2(t) \rangle_{\tau}$, respectively.

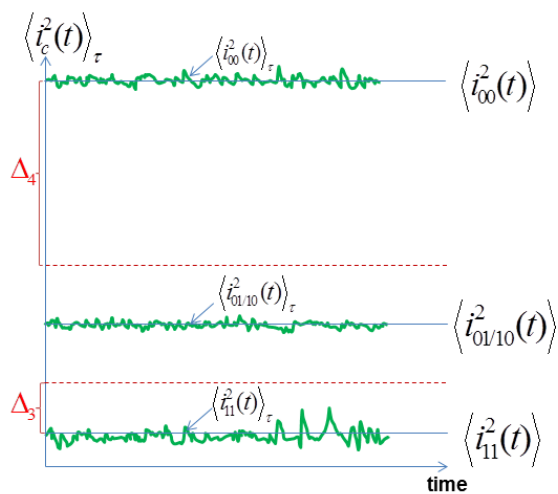

**Fig. 2** Illustration of statistical fluctuations of the finite-time mean-square current levels around their mean values for the 11, 01/10 and 00 bit situations. The scales are arbitrary. Solid lines denote exact (infinite) time average results. $\Delta_3$ and $\Delta_4$ are thresholds for bit interpretation.

Thresholds determine the boundaries between the different interpretations of the measured mean-square channel voltages [31]. In the present paper, we use threshold values $\Delta_3$ and $\Delta_4$ to interpret the measured mean-square channel current over the time window $\tau$, as indicated in Fig. 2. The interpretation is 11 when $\langle i_c^2(t) \rangle_{\tau} < \langle i_{11}^2(t) \rangle + \Delta_3$, and 00 when $\langle i_c^2(t) \rangle_{\tau} > \langle i_{00}^2(t) \rangle - \Delta_4$, respectively. The secure case 01/10 is interpreted as such when $\langle i_{11}^2(t) \rangle + \Delta_3 \leq \langle i_c^2(t) \rangle_{\tau} \leq \langle i_{00}^2(t) \rangle - \Delta_4$. In earlier work [31], the corresponding voltage-based threshold values $\Delta_1$ and $\Delta_2$ were chosen, for normalization purposes, to be proportional to the related mean-square voltages, namely,

$\Delta_1 = \beta \langle Du_{00}^2(t) \rangle$ with $0 < \beta < 1$ and
$\Delta_2 = \delta \langle Du_{11}^2(t) \rangle$ with $0 < \delta < 1$, for the bit situations 00 and 11, respectively. We choose $\Delta_3$ and $\Delta_4$ in a similar way below.

## 3 Error probabilities due to statistical inaccuracies in noise current measurements

Bit errors occur when the protocol makes incorrect bit interpretations due to statistical inaccuracies in the measured mean square noise current, and an error analysis for voltage-based operation was presented before [31]. There are different types of error situations, as shown in Table 1.

**Table 1** Types of errors in the KLJN bit exchange scheme for voltage-based operation [31].

| Measurement Interpretation (Decision) | Actual Situation | | |
|---|---|---|---|
| | 00 | 11 | 01/10 |
| 00 | Correct | Error removed (automatic) | Error removed (automatic) |
| 11 | Error removed (automatic) | Correct | Error removed (automatic) |
| 01/10 | Error * | Error * | Correct |

*The paper addresses these errors and their probability.*

Similarly to the voltage-based case [31], two types of errors need to be addressed for current-based measurements: the 11==>01/10 errors, *i.e.*, errors when the actual situation 11 is interpreted as 01/10, and the 00==>01/10 errors when the actual situation 00 is interpreted as 01/10. The probabilities for these types of errors are estimated below in a similar way as before [31].

3.1 Probability of 11==>01/10 type errors in current-based measurement

We set $R_0 = R$ and $R_1 = \alpha R$, with $\alpha \gg 1$. The mean-square channel noise current for infinite-time average at the 11 bit situation is given by

$$\langle i_{11}^2(t) \rangle = S_{i,11}(f) B_{KLJN}, \qquad (2)$$

where $S_{i,11}(f)$ is the power density spectrum of the channel current at the bit situation 11. From Eq. (1) we obtain

$$\langle i_{11}^2(t) \rangle = 4kT_{eff} \frac{1}{(1+\alpha)R} B_{KLJN}. \qquad (3)$$

Figure 3 shows a block diagram for the measurement process at the 11 bit situation. The channel current first enters a squaring unit. For typical practical applications, the output signal is a voltage, because the squaring unit employs voltage-signal-based electronics. However, for the sake of simplicity and without loosing generality, we assume that the numerical values of the voltage correspond to the measured current. Thus we keep the current-based notation as if the electronics would be a current-based signal system. In other words, the voltages are calibrated so that the numerical values are the same as those of the current. The numerical value of this instantaneous amplitude is expressed as $Qi_{11}^2(t)$, where the constant $Q = \dfrac{1}{Amper}$ denotes the transfer coefficient of the hypothetical multiplier device providing a volt unit also for the square value [37]. This instantaneous amplitude then enters an averaging unit and, after averaging for the finite duration $\tau$, the measurement result is mathematically expressed as $\langle Qi_{11}^2(t) \rangle_\tau = \langle Qi_{11}^2(t) \rangle + i_\tau(t)$, where $i_\tau(t)$ is the AC component remaining after the finite-time average of $Qi_{11}^2(t)$. This averaging process can be represented as low-pass filtering with a cut-off frequency $f_B$ inversely proportional to $\tau$, *i.e.*, $f_B \approx 1/\tau$.

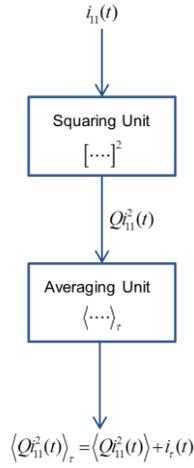

$$\langle Qi_{11}^2(t)\rangle_\tau = \langle Qi_{11}^2(t)\rangle + i_\tau(t)$$

**Fig. 3** Measurement process at bit situation 11. $\langle\ \rangle$ denotes infinite time (exact) average, $Q$ is transfer coefficient of a hypothetical squaring device, and $i_\tau(t)$ is the noise component of the finite-time average of the square of the current.

The AC component $i_\tau(t)$ of the finite-time average is Gaussian, which follows from the Central Limit Theorem because $\tau$ is much larger than the correlation time for the AC component $i_{2,11}(t) = Qi_{11}^2(t) - \langle Qi_{11}^2(t)\rangle$ of $Qi_{11}^2(t)$ since $f_B \square$ . Thus the probability of the 11==>01/10 type of errors is the probability that $i_\tau(t)$ is beyond the threshold, i.e., $i_\tau(t) > \Delta_3$. This probability can be evaluated from the error function, but such a procedure requires numerical integration. However, one can achieve an analytic solution by using Rice's formula [38, 39] of threshold crossings, as discussed next.

Rice's formula can be employed to compute the mean frequency by which $i_\tau(t)$ crosses the threshold value $\Delta_3$ [31]. If we define $S_{i,\tau}(f)$ as the power density spectrum of $i_\tau(t)$, the mean frequency of level crossing can be expressed as

$$\nu(\Delta_3) = \frac{2}{\hat{i}_\tau}\exp\left(\frac{-\Delta_3^2}{2\hat{i}_\tau^2}\right)\sqrt{\int_0^\infty f^2 S_{i,\tau}(f)df}\ , \quad (4)$$

where $\hat{i}_\tau$ denotes the RMS value of $i_\tau(t)$ and is given by $\hat{i}_\tau = \sqrt{\langle i_\tau^2(t)\rangle} = \sqrt{\int_0^\infty S_{i,\tau}(f)df}$. For normalization purposes, we define the threshold value $\Delta_3$ as a fraction of the measured mean-square channel noise current, i.e.,

$$\Delta_3 = \lambda\langle Qi_{11}^2(t)\rangle = \lambda QS_{i,11}(f)B_{KLJN}\ ,\ \text{for}\ 0 < \lambda < 1 \quad (5)$$

The power spectral density $S_{i,2,11}(f)$ for the AC component $i_{2,11}(t)$ of $i_{11}^2(t)$ is considered next. According to previous work [31, 37], and also as given in Fig. 4, $S_{i,2,11}(f)$ can be written

$$S_{i,2,11}(f) = 2Q^2 B_{KLJN} S_{i,11}^2(f)\left(1 - \frac{f}{2B_{KLJN}}\right),$$

$$\text{for}\quad 0 \le f \le 2B_{KLJN}, \quad (6)$$

and $S_{i,2,11}(f) = 0$ otherwise. The low-pass filtering effect of the time averaging cuts off this spectrum for $f > f_B$ but keeps the $S_{i,2,11}(f)$ spectrum for $f < f_B$. Considering that $f_B \ll B_{KLJN}$, the value of $S_{i,2,11}(f)$ can be approximated by its maximum, i.e., $S_{i,\tau}(f) \approx S_{i,2,11}(0)$. Setting $\gamma = B_{KLJN}/f_B$, one obtains

$$\hat{i}_\tau = \sqrt{\int_0^\infty S_{i,\tau}(f)df} \approx \sqrt{f_B S_{i,2,11}(0)} = \sqrt{2Q^2\gamma f_B^2 S_{i,11}^2(f)} \quad (7)$$

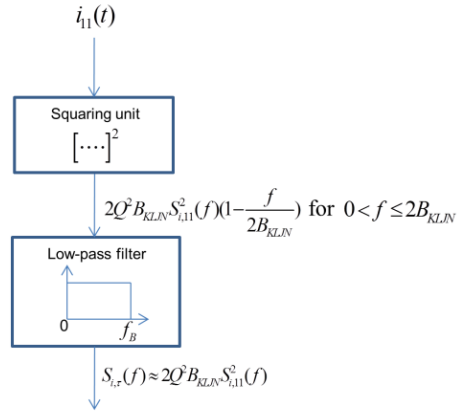

**Fig. 4** Spectra at bit situation 11. $f_B$ is the cut-off frequency for low-pass filtering, $S_{i,11}(f)$ and $S_{i,\tau}(f)$ are power density spectra of the channel current at the bit

situation 11 and of the noise component $i_\tau(t)$, respectively.

The frequency for unidirectional level crossings $\nu(\Delta_3)$, which is half of the value given by Rice's formula, is

$$\nu_\uparrow(\Delta_3) = \frac{1}{\hat{i}_\tau} \exp\left(\frac{-\Delta_3^2}{2\hat{i}_\tau^2}\right) \sqrt{\int_0^\infty f^2 S_{i,\tau}(f)df} \quad , \quad (8)$$

where

$$\Delta_3 = \lambda Q S_{i,11}(f)\gamma f_B. \quad (9)$$

Using Eqs. (7) and (9) one obtains

$$\nu_\uparrow(\Delta_3) = \frac{f_B}{\sqrt{3}} \exp\left(\frac{-\lambda^2 \gamma}{4}\right). \quad (10)$$

Thus the probability $\varepsilon_{i,11}$ of the 11==>01/10 type of errors is

$$\varepsilon_{i,11} \approx \nu_\uparrow(\Delta_3)\tau \approx \frac{\nu_\uparrow(\Delta_3)}{f_B} = \frac{1}{\sqrt{3}} \exp\left(\frac{-\lambda^2 \gamma}{4}\right). \quad (11)$$

It should be noted that this error probability is an exponential function of the parameters $\lambda$ and $\gamma$, which is consistent with earlier results [31]. The dependence on $\gamma$ shows that the error probability decays exponentially with increasing bit exchange period $\tau$.

### 3.2 Probability of 00==>01/10 type errors in current-based measurements

In order to compute this probability, we introduce $\rho$ to define the threshold $\Delta_4$ as a fraction of the measured mean-square channel noise. Thus

$$\Delta_4 = \rho\langle Qi_{00}^2(t)\rangle = \rho Q S_{i,00}(f) B_{KLJN},$$
$$\text{for } 0 < \rho < 1, \quad (12)$$

where $S_{i,00}(f)$ is the channel noise spectrum at the 00 bit situation.

Following the same procedure as above, the probability $\varepsilon_{i,00}$ of the 00==>01/10 type of errors is again found to be exponentially scaling according to

$$\varepsilon_{i,00} = \frac{\nu(\Delta_4)}{f_B} = \frac{1}{\sqrt{3}} \exp\left(\frac{-\rho^2 \gamma}{4}\right), \text{ for } 0 < \rho < 1. \quad (13)$$

### 3.3 Illustration of results with practical parameters

Setting $\gamma = 100$ and $\lambda = 0.5$, the probability $\varepsilon_{i,11}$ for 11==>01/10 type of errors is

$$\varepsilon_{i,11} = \frac{1}{\sqrt{3}} \exp\left(\frac{-\lambda^2 \gamma}{4}\right) \approx 0.001 \quad . \quad (14)$$

Increasing the parameter $\gamma$, and consequently $\tau$, by a factor of two reduces the error probability to $\varepsilon_{i,11} \approx 10^{-6}$.

The bit error probability $\varepsilon_{i,00}$ for the 00==>01/10 type of errors can be computed analogously to the bit error probability $\varepsilon_{i,11}$. In our case of $\alpha \gg 1$, the mean square noise level at 11 is much closer to the value at 01/10 than to the value at 00 (*cf.*, Fig. 2 as an illustration). Therefore, the bit error probability $\varepsilon_{i,00}$ will be significantly smaller than the bit error probability $\varepsilon_{i,11}$. This situation is the opposite for the case of the voltage-based method [31]. Accordingly the experimental test of the KLJN scheme [30] used either the voltage or the current data for decision, depending of which scheme gave the smaller bit error probability.

## 4 An effective error removal method

Below we show a new error removal strategy, utilizing both voltage and current measurements without applying any error correction algorithm, which is superior to the method used in earlier work [30].

Let us assume that Alice and Bob measure both $\langle u_c^2 \rangle_\tau$ and $\langle i_c^2 \rangle_\tau$. In an ideal error-free situation, the same bit interpretations ensue from both mean-square channel noise amplitudes. However, the bit interpretations can differ when there are errors, because the current and voltage amplitudes are statistically independent due to the Second Law of Thermodynamics (*cf.* Eq. 6) and the Gaussian nature of the noises (when the crosscorrelation between two Gaussian processes with zero mean is zero, the two processes are statistically independent). To eliminate errors, we select the cumulative measurement output that has the smallest error associated with it; see Fig. 5 and Table 2. We make use of the fact that, in the bit situation when the current evaluation method has maximum error probability, the voltage-based method has minimum error probability, and *vice versa*. Figure 5 shows the three possible mean-square channel noise current and voltage levels. The threshold values $\Delta_1$, $\Delta_2$, $\Delta_3$ and $\Delta_4$ again provide the boundaries for interpreting the measured mean-square voltage and current values.

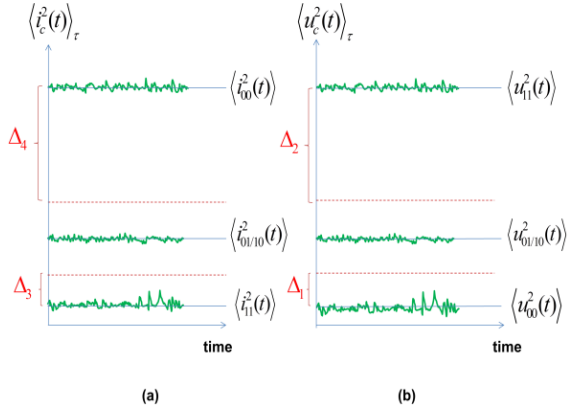

**Fig. 5** Mean-square channel noise measurements of current (a) and voltage (b). ($\Delta_1$, $\Delta_2$) and ($\Delta_3$, $\Delta_4$) are the thresholds for interpreting the measured mean-square voltage and current values, respectively. ($\langle i_{11}^2(t)\rangle_\tau$, $\langle i_{01/10}^2(t)\rangle_\tau$, $\langle i_{00}^2(t)\rangle_\tau$) and (, $\langle u_{11}^2(t)\rangle_\tau$ $\langle u_{01/10}^2(t)\rangle_\tau$, $\langle u_{00}^2(t)\rangle_\tau$) are the mean-square channel noise current and voltage at the 11, 01/10 and 00 bit situations, respectively.

**Table 2** KLJN error removal method with combined current and voltage analysis.

| | | Voltage measurement interpretation | | |
|---|---|---|---|---|
| | | 00 | 11 | 01/10 |
| Current measurement interpretation | 00 | 00 (Insecure/ Discard) | Discard (check attack) | 00 (Insecure/ Discard) |
| | 11 | Discard (check attack) | 11 (Insecure/ Discard) | 11 (Insecure/ Discard) |
| | 01/10 | 00 (Insecure/ Discard) | 11 (Insecure/ Discard) | 01/10 (Secure) |

The only output that is kept is when *both* the current and voltage bit interpretations are secure, *i.e.*, when both are 01/10. For instance, suppose that the bit interpretation obtained from the current measurement is 00 and that the bit interpretation for the voltage measurement is 01/10. In this case, we assume 00 as the correct bit interpretation and hence discard the bit.

**5 Error probabilities in the combined current–voltage analysis method**

The current and voltage noises are independent as a consequence of the Second Law of Thermodynamics and the Gaussianity of thermal noise, [27, 29], and hence the probability of errors in the combined current–voltage analysis method is the product of the error probabilities of the current-based and voltage-based methods.

**5.1 Probability of 00==>01/10 type errors in combined current–voltage analysis**

The probability $\varepsilon_{00}$ of the 00==>01/10 type of errors in the voltage-based method is $\varepsilon_{00} = \frac{1}{\sqrt{3}}\exp\left(\frac{-\beta^2\gamma}{4}\right)$ for $0 < \beta < 1$, as reported before [31], and the probability $\varepsilon_{i,00}$ of the 00==>01/10 type of errors in the current-based method is $\varepsilon_{i,00} = \frac{1}{\sqrt{3}}\exp\left(\frac{-\rho^2\gamma}{4}\right)$ for $0 < \rho < 1$, as shown above. Thus the probability $\varepsilon_{t,00}$ of the 00==>01/10 type of errors in the combined method is given by

$$\varepsilon_{t,00} = \varepsilon_{00}\varepsilon_{i,00} = \frac{1}{3}\exp\left(\frac{-\gamma(\beta^2+\rho^2)}{4}\right),$$

for $0 < \beta < 1$ and $0 < \rho < 1$. (15)

This error probability is again an exponential function of the parameters.

**5.2 Probability of 11==>01/10 type errors in combined current–voltage analysis**

By following the same procedure as above, we find that the probability $\varepsilon_{t,11}$ of the 11==>01/10 type of errors in the combined voltage and current measurements is also exponential

$$\varepsilon_{t,11} = \varepsilon_{11}\varepsilon_{i,11} = \frac{1}{3}\exp\left(\frac{-\gamma(\delta^2 + \lambda^2)}{4}\right),$$

for $0 < \delta < 1$ and $0 < \lambda < 1$. (16)

5.3 Illustration of results with practical parameters

To demonstrate the results for the bit error probability, we assign practical values to the parameters $\beta$, $\rho$ and $\gamma$. For $\gamma = 100$ and $\beta = \rho = 0.5$, we find that $\varepsilon_{t,00}$ is

$$\varepsilon_{t,00} = \varepsilon_{00}\varepsilon_{i,00} = \frac{1}{3}\exp\left(\frac{-\gamma(\beta^2 + \rho^2)}{4}\right) = 1.24 \times 10^{-6}.$$

(17)

If the duration of the bit exchange period, *i.e.*, $\gamma$, is increased by a factor of two (meaning that the speed is decreased by the same factor), the total bit error probability $\varepsilon_{t,00}$ is decreased to $\varepsilon_{t,00} \approx 4.6 \times 10^{-12}$.

6  Conclusion and final remarks

We classified and evaluated the types of errors that occur in the current-based scheme of the KLJN key exchange. These error probabilities showed an exponential dependence on the duration of the bit exchange, which is analogous to the result for the corresponding voltage-based scheme as discussed in earlier work [31].

Furthermore, we presented an error mitigation strategy based on the combination of voltage-based and current-based schemes: only those exchanged bits are kept that are indicated to be secure by both the current and voltage methods. The resulting error probability of this combined strategy is the product of the error probabilities of the two methods, which follows from the statistical independence of the current and voltage measurements. As a consequence, the KLJN scheme can operate without error correcting algorithms, thereby preserving the independence of the exchanged bits of the secure key. Thus the key bits remain independently and identically distributed random variables, which is an important advantage for secure communication [27].

Acknowledgments

Discussions with Elias Gonzalez are appreciated. Y. Saez is grateful to IFARHU/SENACYT for supporting her PhD studies at Texas A&M. R. Mingesz's contribution is supported by the European Union and the European Social Fund. Project #TÁMOP-4.2.2.A-11/1/KONV-2012-0073.

References

1. Liang, Y., Poor, H.V., Shamai, S.: Information theoretic security. Foundations Trends Commun. Inform. Theory **5**, 355–580 (2008). doi: 10.1561/0100000036
2. Bennett, C.H., Brassard, G., Breidbart, S., Wiesner, S.: Quantum cryptography, or unforgeable subway tokens. Advances in Cryptology: Proceedings of Crypto '82, pp. 267–275, Plenum Press, Santa Barbara (1982)
3. Yuen, H.P.: On the foundations of quantum key distribution – Reply to Renner and beyond (2012). arXiv:1210.2804
4. Yuen, H.P.: Unconditional security in quantum key distributions (2012). arXiv: 1205.5065v2
5. Hirota, O.: Incompleteness and limit of quantum key distribution theory (2012). arXiv:1208.2106v2
6. Renner, R.: Reply to recent skepticism about the foundations of quantum cryptography (2012). arXiv:1209.2423v1
7. Yuen, H.P.: Security significance of the trace distance criterion in quantum key distribution (2012). arXiv:1109.2675v3
8. Merali, Z.: Hackers blind quantum cryptographers. Nature News (29 August 2009). doi:10.1038/news.2010.436
9. Gerhardt, I., Liu, Q., Lamas-Linares, A., Skaar, J., Kurtsiefer, C., Makarov, V.: Full-field implementation of a perfect eavesdropper on a quantum cryptography system. Nature Commun. **2**, 349 (2011). doi:10.1038/ncomms1348
10. Lydersen, L., Wiechers, C., Wittmann, C., Elser, D., Skaar, J., Makarov, V.: Hacking commercial quantum cryptography systems by tailored bright illumination. Nature Photonics **4**, 686–689 (2010). doi:10.1038/NPHOTON.2010.214
11. Gerhardt, I., Liu, Q., Lamas-Linares, A., Skaar, J., Scarani,V., Makarov, V., Kurtsiefer, C.: Experimentally faking the violation of Bell's inequalities. Phys. Rev. Lett. **107**, 170404 (2011). doi:10.1103/PhysRevLett.107.170404

12. Makarov, V., Skaar, J.: Faked states attack using detector efficiency mismatch on SARG04, phasetime, DPSK, and Ekert protocols. Quantum Inform. Comput. **8**, 622–635 (2008)
13. Wiechers, C., Lydersen, L., Wittmann, C., Elser, D., Skaar, J., Marquardt, C., Makarov, V., Leuchs, G.: After-gate attack on a quantum cryptosystem. New J. Phys. **13**, 013043 (2011). doi: 10.1088/1367-2630/13/1/013043
14. Lydersen, L., Wiechers, C., Wittmann, C., Elser, D., Skaar, J., Makarov, V.: Thermal blinding of gated detectors in quantum cryptography. Opt. Express **18**, 27938–27954 (2010). doi: 10.1364/OE.18.027938
15. Jain, N., Wittmann, C., Lydersen, L., Wiechers, C., Elser, D., Marquardt, C., Makarov, V., Leuchs, G.: Device calibration impacts security of quantum key distribution. Phys. Rev. Lett. **107**, 110501 (2011). doi:10.1103/PhysRevLett.107.110501
16. Lydersen, L., Skaar, J., Makarov, V.: Tailored bright illumination attack on distributed-phase-reference protocols. J. Mod. Opt. **58**, 680–685 (2011). doi:10.1080/09500340.2011.565889
17. Lydersen, L., Akhlaghi, M.K., Majedi, A.H., Skaar, J., Makarov, V.: Controlling a superconducting nanowire single-photon detector using tailored bright illumination. New J. Phys. **13**, 113042 (2011). doi:10.1088/1367-2630/13/11/113042
18. Lydersen, L., Makarov, V., Skaar, J.: Comment on "Resilience of gated avalanche photodiodes against bright illumination attacks in quantum cryptography". Appl. Phys. Lett. **99**, 196101 (2011). doi:10.1063/1.3658806
19. Sauge, S., Lydersen, L., Anisimov, A., Skaar, J., Makarov, V.: Controlling an actively-quenched single photon detector with bright light. Opt. Express **19**, 23590–23600 (2011). doi: 10.1364/OE.19.023590
20. Lydersen, L., Jain, N., Wittmann, C., Maroy, O., Skaar, J., Marquardt, C., Makarov, V., Leuchs, G.: Superlinear threshold detectors in quantum cryptography. Phys. Rev. Lett. **84**, 032320 (2011). doi:10.1103/PhysRevA.84.032320
21. Lydersen, L., Wiechers, C., Wittmann, C., Elser, D., Skaar, J., Makarov, V.: Avoiding the blinding attack in QKD: REPLY (COMMENT). Nature Photonics **4**, 800–801 (2010). doi:10.1038/nphoton.2010.278
22. Makarov, V.: Controlling passively quenched single photon detectors by bright light. New J. Phys. **11**, 065003 (2009). doi:10.1088/1367-2630/11/6/065003
23. Yuen, H.P.: Key generation: Foundation and a new quantum approach, IEEE J. Sel. Topics Quantum Electr. **15**, 1630–1645 (2009). doi:10.1109/JSTQE.2009.2025698
24. Salih, H., Li, Z.H., Al-Amri, M., Zubairy, H.: Protocol for direct counterfactual quantum communication. Phys. Rev. Lett. **101**, 170502 (2013). doi:10.1103/PhysRevLett.110.170502
25. Kish, L.B.: Totally secure classical communication utilizing Johnson (-like) noise and Kirchhoff's law. Phys. Lett. A **352**, 178–182 (2006). doi:10.1016/j.physleta.2005.11.062
26. Mingesz, R., Kish, L.B., Gingl, Z., Granqvist, C.G., Wen, H., Peper, F., Eubanks, T., Schmera, G.: Unconditional security by the laws of classical physics. Metrol. Meas. Syst. **20**, 3–16 (2013). doi:10.2478/mms-2013-0001
27. Kish, L.B., Abbott, D., Granqvist, C.G.: Critical analysis of the Bennett–Riedel attack on secure cryptographic key distributions via the Kirchhoff-law–Johnson-noise scheme (2013). viXra:1306.0058, arXiv:1306.653
28. Kish, L.B.: Protection against the man in the middle attack for the Kirchhoff-loop Johnson (-like) -noise cipher and expansion by voltage-based security. Fluct. Noise Lett. **6**, L57–L63 (2005). doi:10.1142/S0219477506003148
29. Kish, L.B.: Enhanced secure key exchange systems based on the Johnson noise scheme. Metrol. Meas. Syst. **20**, 191–204 (2013). doi:10.2478/mms-2013-0017
30. Mingesz, R., Gingl, Z., Kish, L.B.: Johnson (-like) -noise–Kirchhoff-loop based secure classical communicator characteristics, for ranges of two to two thousand kilometers, via model-line, Phys. Lett. A **372**, 978–984 (2008). doi:10.1016/j.physleta.2007.07.086
31. Saez, Y., Kish, L.B.: Errors and their mitigation at the Kirchhoff-law–Johnson-noise secure key exchange (2013). viXra:1305.0126; arXiv:1305.4787v1
32. Kish, L.B., Kwan, C.: Physical uncloneable function hardware keys utilizing Kirchhoff-law–Johnson-noise secure key exchange and noise-based logic (2013). viXra:1305.0068; arXiv:1305.3248
33. Kish, L.B., Saidi, O.: Unconditionally secure computers, algorithms and hardware. Fluct. Noise Lett. **8**, L95–L98 (2008). doi:10.1142/S0219477508004362
34. Gonzalez, E., Kish, L.B., Balog, R., Enjeti, P.: Information theoretically secure, enhanced Johnson noise based key distribution over the smart grid with switched filters, PLoS ONE **8**, e70206 (2013). doi:10.1371/journal.pone.0070206
35. Kish, L.B., Mingesz, R.: Totally secure classical networks with multipoint telecloning (teleportation) of classical bits through loops with Johnson-like noise. Fluct. Noise Lett. **6**, C9–C21 (2006). doi:10.1142/S021947750600332X
36. Kish, L.B., Peper, F.: Information networks secured by the laws of physics. IEICE Trans. Commun. **E95-B**, 1501–1507 (2012)
37. Kish, L.B., Mingesz, R., Gingl, Z., Granqvist, C.G.: Spectra for the product of Gaussian noises. Metrol. Meas. Syst. **19**, 653–658 (2012). doi: 10.2478/v10178-012-0057-0
38. Rice, S.O.: Mathematical analysis of random noise. Bell System Tech. J. **23**, 282–332 (1944). http://archive.org/details/bstj23-3-282
39. Rychlik, I.: On some reliability applications of Rice's formula for the intensity of level crossings. Extremes **3**, 331–348 (2000). doi:10.1023/A:1017942408501